\documentstyle[12pt]{article}
\title{Binding energy in two and three-body relativistic dynamics
\footnote{Talk presented at the Workshop "Critical Stability 
of Few-Body Quantum Systems", Les Houches, Oct.8-12, 2001} }
\author{Ph. Droz-Vincent\\[2mm]DARC/LUTH  \      Observatoire de Meudon\\
5 place Jules Janssen 92195 Meudon, France}

\date{ }

\begin{document}
\maketitle
\abstract{Two-body  and three-body systems of scalar bosons are considered
 in the framework of covariant constraint dynamics.  The reduced equation 
obtained after eliminating redundant degrees of freedom can be viewed as an 
eigenvalue equation for an observable which is intimately related with the 
relative motion. We display the connection of this observable with binding 
energy.}

\medskip


\section{Introduction}
$$                                                 $$
\noindent
In galilean mechanics, the binding energy of a bound state is just the energy 
of relative motion. It is an observable and can be expressed in terms of  
dynamical variables.

\noindent 
In the framework of special relativity, in contrast, the binding energy
of a bound state is usually defined as the mass defect  $ \sum m  - M$ where
 $M$ is the total mass and $\sum m$ refers to the  constituent masses. 
But it is still natural to look for  some observable, depending on the 
relative degrees of freedom, that could precisely characterize the relative
 motion of the system.
Of course, in case we succeed, a comparison of this quantity with 
       $ \sum m  - M$ is desirable.

\noindent
 There are many different formulations of few-body relativistic dynamics 
\cite{barcelo} \cite{carbo}.
Our approach is based upon coupled wave equations \cite{droz2} 
\cite{belstern} that can be interpreted as 
mass-shell  constraints \cite{tod}.  
For scalar particles, these constraints deal with  
{\em squared}-mass  operators that include an interaction term.
In contrast, all the  generators of the Poincar\'e algebra are free of 
interaction.             In this report, we focus on scalar particles.  

\noindent
For two (resp. three)  particles we start from  a pair (resp. a triple)
 of coupled  equations, for one  wave function depending on two (resp. three)
 four-dimensional arguments. 
In this formulation, like in the Bethe-Salpeter (BS) equation,
 covariance is paid by  redundant degrees of freedom.
 
But demanding that the total linear momentum four-vector 
$P= \sum  p$ has a sharp value $k$,
and solving the difference(s)  of the constraint equations,  
it is possible to factorize out one (resp. two) degrees of freedom, and one is
left with a reduced wave equation, in the form of an eigenvalue problem; 
solving this problem determines  the   total mass   $M  = \sqrt{k \cdot k}  $.

\noindent
This procedure is straightforward for two-body systems  \cite{droz2},
 but in the  three-body case, it  requires a nontrivial transformation
\cite{droz3}.
                          
\medskip
\noindent
In both cases, by an elementary combination of the mass-shell constraints, one 
can exhibit a constant of the motion, say    $ N = - \Lambda $
which is homogeneous to a squared mass,
depends only on relative variables (therefore is translation invariant),
  includes the  term of mutual interaction,  and  is   intimately related with
 relative motion. Our purpose is to clarify the relationship between this 
observable and the mass defect.
    
\noindent It seems  natural to define slow relative motion 
 by the condition that the eigenvalue $\lambda$ of   $\Lambda$ 
 is small (in absolute value) with respect to the squares  of the constituent
 masses, $m_a$ 
(this situation in turn  implies that also  $|\lambda |  \ll   M ^ 2$).

\medskip
\noindent   {\bf Notation}

\noindent The dynamical variables associated with particle $a$ are the 
four-dimensional conjugate quantities $q_a ,  p_a$.
 Particle labels are $a,b = 1,2 $ (resp. $a,b = 1, 2, 3$).

\section{Two-Body System}

For the class of models that we consider, the same interaction term $2V$ arises 
in both sqared-mass operators. This scheme seems  to accomodate practically 
all realistic interactions inspired from quantum field theory. Its 
relationship with either the quasi-potential approach \cite{todbarcelo} 
\cite{todsazriz}  or the BS  equation    \cite{sazBS}  has 
been demonstrated, and fermionic generalizations have been elaborated.
\noindent We introduce  relative variables; they are the four-vectors 
\begin{equation}  
 z= q_1 - q_2 ,     \qquad  \      y = {1 \over 2} (p_1 - p_2)  \end{equation}
We have the transverse parts 
\begin{equation}   
  {\widetilde z} = z - {z \cdot P \over P^2 }  P ,    \qquad
           {\widetilde y} = y - {y \cdot P \over P^2 }  P ,    \qquad
                                                            \end{equation}
For arbitrary masses it is convenient to set
\begin{equation}     \mu = {1 \over 2} (m_1 ^2 +  m_2 ^2 ),     \qquad \quad
  \nu = {1 \over 2} (m_1 ^2 -  m_2 ^2 )         \label{defnu}   \end{equation}
The reduced wave equation takes on the form
\begin{equation}    (N+ \lambda) \phi = 0  
                                            \label{redu}         \end{equation}
with
 \begin{equation}    
 N  =  { {\widetilde y} }^2  +  2 V   \label{defN2}          \end{equation}
\begin{equation}    \lambda =
{M^2 \over 4} + {\nu ^ 2 \over M^2} -  \mu               
                                              \label{lambda}      \end{equation}
The interaction term  $V$ has the dimension of a   squared mass. It may 
essentially depend on $ {{\widetilde z}} ^2$  and  $P^2$ (possibly also on 
${\widetilde y} ^2, 
\    {\widetilde z} \cdot {\widetilde y} ,\    y \cdot P $).  
 On shell and using  the rest frame, we simply have 
                  $ {{\widetilde z}} ^2 =   -{\bf z}^2        $.
Therefore  (\ref{redu}) can be identified with a nonrelativistic equation 
\cite{martin}, 
at least when the "potential"  $V$ doesnot depend on $P^2$. 
But realistic interactions may bear  some dependence on $P^2$ 
(energy dependent potentials); in this case, (\ref{redu}) becomes   
 a nonconventional eigenvalue problem. This  complication has been 
discussed in the literature \cite{todsazriz} \cite{sazBS}.

\medskip
\noindent      {\bf Weak Binding}

\noindent  Eq. (\ref{lambda}) can be solved for  $M^2$.
 Insofar as  $\nu$ is not too large, we obtain 
\begin{equation}   M^2 =
 2 (\lambda + \mu ) + 2 \sqrt{ (\lambda + \mu)^2  -  \nu ^2 }  \end{equation}   
For  slow relative motion,  $|\lambda |  \ll  m_a^2 $,
 we can develop $M^2$ in powers   of  $\lambda $.
According to (\ref{defnu}) we get
$$ M^2 =  (m_1 + m_2 )^2   + \lambda \    { (m_1 + m_2 )^2   \over  m_1 m_2 } 
                   +  O  (\lambda ^2)                                    $$
hence
\begin{equation}  m_1 + m_2  -  M = 
  -  {\lambda \over 2 m_0 } + 
                        \cdots      \label{bind2}         \end{equation}
It is clear that slow relative motion corresponds to weak binding.

\section{Three-Body System}
\noindent   The squared mass operators are $p_a^2 + 2V $.
 This  formulation  aims at the  elimination of two superfluous 
degrees of freedom.                         

\noindent
For three-body systems it is difficult to find an interaction such that the  
 mass-shell constraints are compatible among themselves,
 respect Poincar\'e invariance, reduce to three Klein-Gordon equations in the
 absence of interaction, and allow for 
eliminating two redundant degrees of freedom.
All these requirements can be 
satisfied however, in a tractable manner, 
 with help of a  "point transformation in momentum space".
In order to formulate this transformation, it is essential to introduce 
relative variables  as follows \cite{saz3}.

\noindent   Relative-particle indices are   $A, B = 2, 3 $.
We define   the four-vectors
\begin{equation}
 z_A = q_1 - q_A  ,  \qquad \qquad  y_B =  {P \over 3} -  p _B       
\label{relative}    \end{equation}   
The transverse part are 
 ${\widetilde z} _A , {\widetilde y} _B$ and ${\widehat z} , {\widehat y} $ 
respectively with respect tp $P$ and $k$. In the rest frame we have
$         {{\widehat y} }^2 _A =  - {\bf y}^2 _A  , 
 \quad    {{\widehat z} }^2 _A =  - {\bf z}^2 _A  $, etc.

\noindent
Our transformation \cite{droz3} is characterized by
\begin{equation}     (p_1 + p_A)\cdot (p_1 - p_A) = P \cdot  (p'_1 -p'_A)  
\label{transfo}     \end{equation}
and by the requirement that it leaves $P$ and 
${\widetilde y_A}$ invariant 
(the new relative momenta are of course  $y'_A = P/3 - p' _A$  ).

This procedure  generates a canonical transformation, giving rise 
to  new configuration variables   $z'_A$.
The difference equations are mapped to 
$  y'_A \cdot P = c_A  \Psi $
where the constants $c_A$ are combinations of the squared constituent
  masses; the dependence of $\Psi $ on the {\em new} relative times is
 factorized out.

Here, for  simplicity, we assume  that  $m_a = m$.

\noindent The  sum of the  mass-shell constraints yields the
 reduced equation
\begin{equation} 
  (9 m^2 - M^2 ) \psi =  6 N  \psi    \label{redeq3}        \end{equation}
 where now
\begin{equation}  N = - \Lambda =
{{{\widetilde y}} _2} ^2   +  {{\widetilde y}} _3 ^2 
  + {\widetilde y} _2 \cdot {\widetilde y} _3 + 3 V +  
      P ^2 \Xi                                  \label{defN3}   \end{equation}
In the momentum representation, the reduced wave function $\psi$ depends 
only on ${\widehat y} _2 , {\widehat y} _3$.
In view of the  compatibility requirement, a  closed form of the interaction
is available only in 
terms of the {\em new variables}. A typical example  would be a function 
of ${{\widehat z}}'_2 , {{\widehat z}}' _3 , P^2$.

\noindent
The quantity   $P^2 \Xi$ has no counterpart in two-body systems. Here it 
stems  from having added {\em three} constraints. 
Its exact expression in terms of the new variables amounts to solve a 
fourth-degree algebraic equation and  would be extremely 
complicated (see details in \cite{droz3}). 
Fortunately, it can be naturally expanded in powers of $1/P^2$, 
implying on the mass shell  an expansion in powers of $1/M^2$.
This makes our model more tractable when the constituent particles are light 
with respect to the total mass  of the system.

\medskip
\noindent
When the constituent masses (although different from zero) are small  with
 respect to the total mass $M$, that is  $m^2  \ll M^2$,  
  we can drop the last term in (\ref{defN3}).

\noindent
With  this truncation,  (\ref{redeq3}) is similar to a nonrelativistic 
equation, except for eventual  complications resulting from a possible 
dependence of $N$ on $P^2$.
Moreover, in this limit 
 ${{\widehat z}}'_A $ differs very little from
  ${{\widehat z}}'_A$, which allows for a
weak form of cluster separability \cite{droz3}.

\noindent But when $m^2 / M^2$ is not small enough, 
the on-shell expression of $P^2 \Xi$ must be written as a Taylor expansion 
including several  powers of $1/M^2$, say
$  \displaystyle    M^2  \underline{\Xi} =  {1\over M^2} \Gamma     $,
where $\Gamma$ is regular for  $M \rightarrow \infty$,  say
$ \Gamma = \Gamma _{(0)} + O(1/M^6) $ where
\begin{equation}     \Gamma _{(0)} =
({{\widehat y} _2}^2 )^2 +  ({{\widehat y} _3}^2 )^2 + 
 ({\widehat y} _2 \cdot {\widehat y} _3 )^2
+ ({ {\widehat y} _2}^2  + 
 { {\widehat y} _3}^2 ) \      ( {\widehat y} _ 2 \cdot {\widehat y} _3)
 - {{\widehat y} _2 }^2  {{\widehat y} _ 3 }^2                \end{equation}
Note that $\Gamma$ is a positive operator and would survive in the absence of
interaction.    Irrespective of the shape of the potential, we expect that
it  provides  a positive correction to the truncated expression of  $N$.
A rigorous statement, however, would require solving a nonconventional
eigenvalue problem.

\noindent    At first  order in  $1/M^2$ we have,  in the rest frame
$$ N  \psi =
 ( -{\bf y} ^2 _2 - {\bf y} ^2 _3 -{\bf y } _2 \cdot {\bf y  _3}   +  3V  +
  {\Gamma_{(0)}   \over  M^2} )  \psi   $$
In the rest-frame $\Gamma _ {(0)}$  is bi-quadratic in $\bf y$.

\noindent                 Let us evaluate  the  binding energy,  in
     the general case described by  (\ref{redeq3}). 
Equation  (\ref{redeq3}) can be viewed as a (generalized)  eigenvalue
 equation for 
 $\Lambda$ with eigenvalue $\lambda$, if we set   
$6 \lambda =  M^2 - 9 m^2$. The mass defect is exactly 
\begin{equation}
 \sum m - M =     3m -  \sqrt{9m^2 + 6 \lambda}        \end{equation}
 It would increase together  with the eigenvalue of $N$.

\medskip
\noindent   {\bf Weak  Binding}

\noindent
It is noteworthy that  weakly bound systems are not eligible for the
light-constituent approximation.
Indeed  they have 
$m^2$ almost one order of magnitude smaller than $M^2$,
but this is not enough for dropping terms like $O(1/M^2)$.

\noindent
Assuming that $ |\lambda |  \ll  m^2$ wee can develop the mass defect and find
\begin{equation} 3m - M =
  - {\lambda \over m} + {1 \over 6} {\lambda ^2 \over m ^3}
  + \cdots               \label{bind3}                      \end{equation}
Since we consider equal masses, then $m = 2m_0$ where $m_0$ is 
the reduced mass of either of particles 2, 3, with respect to particle 1. 
Note the analogy of (\ref{bind3}) with (\ref{bind2}).

\section{Conclusion}

The constraint formulation of relativistic two-body dynamics admits 
 a well-  understood  contact with field theory,
  whereas constraint three-particle 
dynamics is still a field of recent investigation. 
    Nevertheless, it is possible to present in parallel ways, for both cases,  
the relationship of binding energy with a remarkable observable $N$ which 
naturally arises in the reduced wave equation.
This situation results from the fact that, in both cases, our basic equations 
involve a unique interaction term and are tailored for allowing  
elimination of the redundant variable(s) implied by manifest covariance.

\noindent
  Binding energy, defined as  the mass defect, has a simple relationship 
with the  eigenvalue of   $N$.
                                    In the case of weak binding, 
these quantities become proportional through  the constant factor $1/2m_0$. 

\noindent
The reduced wave equation can be 
compared and, to some extent, identified  with a  nonrelativistic equation:
in a straightforward manner for two particles (with arbitrary masses),
but  only in the light-constituent  limit for three particles 
(with equal masses); in three-body systems,  weak binding {\it doesnot}
 correspond to  a nonrelativistic  form of the wave equation. 

\medskip
\noindent
For applications, we plan to introduce spin
 and to improve the  contact with other approaches \cite{bij}.

\end{document}